# Correlation between superfluid density and transition temperature in infinite-layer nickelate superconductor Nd$_{1-x}$Sr$_x$NiO$_2$

Z. J. Li[1,*], R. Z. Zhang[1,*,†], M. H. Xu[2,*], K. Y. Liang[1,*], Y. Zhao[2], Q. S. He[1], Q. Z. Zhou[1], B. R. Chen[1], P. H. Zhang[1], K. Z. Yao[1], H. X. Yao[1], L. Qiao[2,†], Y. H. Wang[1,3,†]

1. *State Key Laboratory of Surface Physics and Department of Physics, Fudan University, Shanghai 200433, China*
2. *School of Physics, University of Electronic Science and Technology of China, Chengdu 610054, China*
3. *Shanghai Research Center for Quantum Sciences, Shanghai 201315, China*

\* These authors contributed equally to this work.
† Email address: ruozhouzhang@fudan.edu.cn; liang.qiao@uestc.edu.cn; wangyhv@fudan.edu.cn;

**Abstract**

**A strong correlation between zero-temperature superfluid density ($\rho_{s0}$) and transition temperature ($T_c$) is considered as a hallmark of unconventional superconductivity. However, their relationship has yet to be unveiled in nickelates due to sample inhomogeneity. Here we perform local susceptometry on an infinite-layer nickelate superconductor Nd$_{0.8}$Sr$_{0.2}$NiO$_2$. The sample shows inhomogeneous superfluid density and $T_c$ on micron-scale. The spatial statistics for different scan areas reveal a linear dependence of local $T_c$ on $\rho_{s0}$ for $T_c > 8$ K and a sub-linear one for $T_c < 8$ K. Remarkably, the overall relationship is reminiscent of that reported in overdoped cuprate superconductors, hinting at a close connection between them.**

In the quest to understand the microscopic origin of unconventional superconductivity, a concerted effort has been made to explore the limiting factor of superconducting transition temperature ($T_c$) [1]. In contrast to the Bardeen-Cooper-Schrieffer (BCS) superconductors where $T_c$ is determined by the pairing strength [2], the $T_c$ of unconventional superconductors is always correlated with the zero-temperature superfluid density ($\rho_{s0} \equiv \lambda^{-2} \propto \frac{n_s}{m^*}$, with $\lambda$ the London penetration depth, $n_s$ the superconducting carrier density and $m^*$ the effective electron mass). The latter encodes the stiffness of quantum-mechanical phase of Cooper pairs, reflecting the resilience of the superconductor to thermal or quantum phase fluctuations [3–6]. In hole-doped cuprate superconductors, for instance, the Uemura law, i.e., $T_c \propto \rho_{s0}$, holds in the moderately underdoped regime, which was attributed to strong thermal phase fluctuations [3,7]. As the critical doping where $T_c = 0$ K is approached, a crossover from $T_c \propto \rho_{s0}$ to $T_c \propto \rho_{s0}^\alpha$ [$\alpha = z/(z + D - 2)$ with $z$ the quantum dynamic exponent and $D$ the spatial dimensionality] was observed [8–10], providing the evidence for quantum phase fluctuations in strongly underdoped cuprates [11,12]. In the overdoped regime, previous superfluid density measurements on overdoped La$_{2-x}$Sr$_x$CuO$_4$ (LSCO) films demonstrate that the relation between $T_c$ and $\rho_{s0}$ is generally linear with an offset as doping varies, i.e., $T_c = T_0 + \alpha \rho_{s0}$ ($T_0$, $\alpha > 0$ being constant parameters), but converts into a parabolic scaling, $T_c \propto \rho_{s0}^{1/2}$, for $T_c < 12$ K [13]. The strong correlations between $T_c$ and $\rho_{s0}$ have also been found in heavy-fermion [14], transition metal dichalcogenide [15] and iron-based superconductors [16–18].

It is natural to pose the same question for the recently discovered high-$T_c$ superconductor infinite-layer (IL) nickelates [19–21]. After all, the parent compound of IL nickelates share the same nominal transition metal 3$d^9$ electron configuration and crystal structure as the cuprates. But the answer is not obvious due to some non-trivial differences in the nickelates: the absence of long-range magnetic order in the parent compound [19,22–24], the multiband electronic structure [25,26], and the possible mixture of $s$- and $d$- wave pairing [27]. Whether these differences lead to a different nature in the unconventional superconductivity of nickelates remains to be investigated. To this end, it is important to examine quantitatively the correlation between $T_c$ and $\rho_{s0}$ in nickelates, and compare with that found earlier in cuprates.

The current complexities involved in the material realization of superconducting nickelates make it technically challenging to determine this key relationship. Superconducting IL nickelates $R_{1-x}Sr_xNiO_2$ ($R$ = La, Pr, Nd) are so far synthesized via $CaH_2$-assisted soft-chemistry topotactic reduction on the perovskite $R_{1-x}Sr_xNiO_3$ [19–21]. Such a reduction step is critical for inducing superconductivity in the IL nickelates. However, it also introduces significant lattice distortion and inhomogeneous chemical doping [28–33], which then leads to spatial variation of superconductivity on the micron scale [34]. The inhomogeneity impedes accurate determination of $T_c$ and superfluid density using conventional bulk measurement techniques such as muon spin rotation or two-coil mutual inductance [7,13]. Hence, a highly sensitive local magnetic probe capable of imaging both $T_c$ and superfluid density is essential.

Here, we employ scanning superconducting quantum interference device (sSQUID) microscopy in magnetometry and susceptibility modes to systematically investigate the spatial distribution of $T_c$ and superfluid density in $Nd_{0.8}Sr_{0.2}NiO_2$ (NSNO) films grown on (001)-oriented single-crystal $SrTiO_3$ (STO). Our samples exhibit the relatively high bulk $T_c$ of 8.6 K for NSNO on STO reported so far [19,23,35,36]. Yet, we still find spatial inhomogeneity of $T_c$ and superconducting diamagnetism in different regions of the film. Noticeably there are ring-like patterns showing weaker diamagnetism, which have also been observed previously on NSNO films by different groups [34], proving to be generic features of NSNO/STO. By analyzing the mappings of $T_c$ and susceptibility from different regions on the sample, we further find that the inhomogeneous $T_c$ is spatially correlated with local $\rho_{s0}$: $T_c$ varies linearly with $\rho_{s0}$ for regions with $T_c > 8$ K, but exhibits a sublinear dependence on $\rho_{s0}$ approximate to $T_c \propto \rho_{s0}^{1/2}$ for $T_c < 8$ K. These results are reminiscent of those reported in overdoped LSCO films, strengthening the link between the superconductivity of IL nickelates and that of cuprates.

Two NSNO epitaxial films with thicknesses of 15 nm and 10 nm were grown on STO

(001) substrates using the pulsed laser deposition technique. No capping layer was introduced for both samples. Details of the sample growth and characterization are depicted in Supplementary Materials S1. The overall high-quality (00$l$)-oriented growth of the films was verified by the atomic force microscopy, X-ray diffraction measurements and the scanning transmission electron microscopy (STEM) (Figs. S1-S2). Fig. 1(a) shows the temperature-dependent resistance of one film (the blue line) alongside with the data from previous studies (the brown line) [19,37]. The onset temperature of the superconducting transition is $T_{c,\text{onset}} = 11.5$ K and the bulk zero resistance appears at $T_c = 8.6$ K. The linear-in-temperature resistivity is observed in a wide temperature range of approximately 100 to 300 K, which is consistent with previous reports on optimally doped nickelates [37,38]. It should be noted that the $T_c$ value and the normal-state transport rely on the type of substrate: NSNO grown on (LaAlO$_3$)$_{0.3}$(Sr$_2$TaAlO$_6$)$_{0.7}$ exhibits higher $T_c$ and better conformity of the linear resistivity at low temperatures [37]. Nevertheless, for films grown on the STO substrates, both the $T_c$ and the $T_{c,\text{onset}}$ of this sample are higher and $\frac{\Delta T_c}{T_c} = \frac{T_{c,\text{onset}} - T_c}{T_c}$ is smaller than those reported in refs. [19,23,35,36]. For another NSNO film, it exhibits the highest $T_c$ of 12.6 K (Fig. S10) among NSNO/STO films reported so far (see e.g., refs. [27,39]). These facts indicate the high quality of our NSNO samples.

To characterize the spatial distribution of the superconductivity of the NSNO film, we employ sSQUID magnetometry and susceptometry [40–48], which has high magnetic flux sensitivity even under zero magnetic field (see details in Supplementary Materials S2). Fig. 1(b) shows a large-scale (280 × 150 μm$^2$) susceptometry mapping measured at $T_{\text{base}} = 5.00$ K, which is stitched together from 10 contiguous scans, each with an area of 90 × 90 μm$^2$. The film exhibits an overall diamagnetic background, but contains significant inhomogeneity, consistent with a recent sSQUID study of IL nickelates [34]. The diamagnetism is weaker on the right edge of the image. There are also two weakly diamagnetic rings (WDRs) with sizes of approximately 30 × 50 μm$^2$ (middle) and 20 ×

30 μm² (lower-left). From a zoom-in scan on the largest weakly diamagnetic ring, i.e. the main WDR (indicated by the large arrow), multiple smaller WDRs (indicated by small arrows) can be further seen [Fig. 1(c)]: one close to the main WDR and two small rings located at the top-right corner of the scan area. Since the susceptibility $\chi'$ is directly proportional to the magnitude of superfluid density in a thin film [49], these images indicate spatial variation of superfluid density in the NSNO sample. The inhomogeneous superfluid density can also be found in the second NSNO film with higher $T_c$ (Fig. S10).

We note that for the susceptometry scan, submicron defects could appear as weakly diamagnetic halos due to the geometry of the nano-SQUID sensor layout [50]. Another possible origin is the formation of a mixed Ruddlesden-Popper (RP) secondary phase during the growth process [51], as observed in the STEM cross-sectional images of our sample (Fig. S2). Additionally, loss of oxygen [52] and presence of hydrogen [33] during the reduction process also affect superconductivity. Considering that the $T_c$ of our films is among the highest for NSNO films grown on STO, the observed inhomogeneity may reflect the inherent limitation of the current synthetic method of IL nickelates.

DC magnetic flux image shows weak magnetic-domain like patterns [Fig. 1(d)] [53]. This is different from the strong flux contrast of isolated superconducting vortices [54]. Such patterns do not show obvious spatial correlation to the susceptometry image [Fig. 1(c)], suggesting that their origin is different from that of WDR. Actually, the magnetic-domain like patterns of NSNO have also been reported in ref. [34], and were attributed to NiO$_x$ ferromagnetic nanoparticles in that work. Similar magnetic domains with irregular shapes and typical sizes of 5-20 μm was observed from the ferromagnetic EuO layer of EuO/KTaO$_3$ heterostructure, which coexists with the diamagnetism of the interfacial superconductivity [55]. However, the superfluid density in the NSNO film is relatively large so that the overall susceptibility is still diamagnetic. The susceptibility approach curves $\chi'(z)$ (with $z$ the distance between the nano-SQUID tip and the sample) on selected points around the main WDR [Fig. 1(e)] show that the main WDR is also diamagnetic overall. Notably, the value of $\chi'(z)$ on points inside the main WDR (points 1 and 2) decrease slower than those outside the main WDR (points 3 and 4) as $z$ decreases. $T_c$ is positively correlated with the diamagnetic strength for these four points, which will be demonstrated in the following.

The temperature evolution of the susceptometry images over the same scan area of Fig. 1(c) are shown in Fig. 2. The diamagnetism fades out quickly as the temperature rises from $T = 5.00$ to $8.75$ K above the bulk $T_c$ [Figs. 2(a)-(d)]. The average value of $\chi'$ at $T = 8.75$ K is only half of that at $T = 5.00$ K. Moreover, the diamagnetism of the main WDR and upper corners vanishes faster than that of the lower-right region. In the temperature regime of $T = 8.84$ to $9.63$ K which is around bulk $T_c$, the non-uniform susceptibility evolution is more obvious. As the temperature increases, the main WDR region becomes non-superconducting ($\chi' \sim 0$) first, while the superconductivity in the lower-right region is more persistent, leading to a diamagnetic patch up to $T = 9.09$ K [Figs. 2(e)-(g)]. A few segregated islands remain diamagnetic above $T = 9.33$ K [see e.g., the blue spots near the center of Figs. 2(i)-(k)]. Superconducting diamagnetic patterns disappear completely for $T > 9.63$ K (not shown in the figure). Fig. 2(l) shows the local $T_c$ map of the scan area, which is determined as the onset temperature of superconducting diamagnetism ($\chi' < 0$). The local $T_c$ is spatially inhomogeneous and the lower values appear mainly inside the WDRs. Again, since no correlation is found between the magnetic contrast (Fig. S4) and the diamagnetic patterns in the susceptibility images [Figs. 2(a)-(k)], ferromagnetic clusters are not affecting local diamagnetism in our samples [34].

To quantify the temperature evolution of inhomogeneity, we calculate the autocorrelation plots of the susceptibility maps at different temperatures (Fig. S5). The typical size of the superconducting patches with constant diamagnetic susceptibilities, $l_p$, can be evaluated as the full-width-at-half-maximum of the central peak of the radial profile of the autocorrelation plots (see details in Supplementary Material S4). Similar procedure has been adopted to analysis the gap maps of granular superconductors [56]. As temperature is increased across the bulk $T_c$ of 8.6 K, $l_p$ decreases from a nearly $T$-independent value (~ 13 μm) to a small but finite value (~ 1 μm) [Fig. S6(a)]. The finite value of $l_p$ above bulk $T_c$ is in sharp contrast to the disappearance of $l_p$ at bulk $T_c$ for a homogeneous superconductor. The former is due to the segregated superconducting patches at high temperatures, corroborating the inhomogeneous $T_c$ in NSNO. Meanwhile, we find that the temperature dependence of pararesistance above bulk $T_c$ can be well captured by the formula of zero-dimensional Aslamasov-Larkin fluctuations [57], which again indicates the existence of small-sized superconducting patches at high temperatures [Fig. S6(b)] . All these analyses are consistent with the observed inhomogeneous $T_c$ in NSNO [Fig. 2(l)].

The spatial inhomogeneity of the NSNO film enables us to investigate the correlation between $T_c$ and superfluid density in a statistical way. For a superconducting thin film, the local superfluid density is directly proportional to the diamagnetic susceptibility according to the thin diamagnet film model [49]:

$$\frac{\chi'(z)}{\chi_s} = -\frac{a}{\Lambda}\left(1 - \frac{2\frac{z-z_0}{a}}{\sqrt{1+4(\frac{z-z_0}{a})^2}}\right), \quad (1)$$

where $\chi_s$ is the SQUID self-susceptibility, $a$ is the effective radius of the field coil, $z_0$ is the offset of sample-coil distance, and $\Lambda$ is the Pearl length, which is proportional to the inverse of superfluid density ($\Lambda^{-1} = \frac{d}{2}\lambda^{-2}$ with $d$ being the film thickness). Figs. 3(a)-(f) depict the susceptibility approach data (the dots) and the fitting curves (the solid lines) of Eq. (1) for selected points. We note that although the low-temperature approach data can be well fitted with Eq. (1), the fitting curves exhibit non-negligible deviations from the experimental data at some temperatures near $T_c$. The relatively poor fitting of the approach data to Eq.(1) near $T_c$ was also found by ref.

[34] , and was attributed to the strong paramagnetic response at relatively high temperatures. Besides, the significant spatial inhomogeneity of superfluid density near $T_c$, with a length scale smaller than the Pearl length, could lead to the breakdown of Eq. (1) that assumes a homogeneous superfluid density [58].

Fig. 3(g) shows the extracted temperature-dependent superfluid density $\Lambda^{-1}(T)$ from the susceptibility approach curves. We observe nearly $T$-linear behavior of $\Lambda^{-1}$ for $T < 7$ K on all points, consistent with the results reported in ref. [34]. The relatively rapid drop of superfluid density near $T_c$ is reminiscent of the Berezinskii–Kosterlitz–Thouless (BKT) transition in 2D superconductors originated from the unbinding of vortex-antivortex pair by thermal phase fluctuations [59,60]. We further fit the $\Lambda^{-1}(T)$ relations with a phenomenological formula [see the solid lines in Fig. 3(g)]:

$$\Lambda^{-1} = \Lambda_0^{-1}(1 - \frac{T}{T_c})^\alpha, \qquad (2)$$

where $\Lambda_0^{-1} = \Lambda^{-1}(0\ \text{K})$, and $\alpha$ is a parameter that possibly relates to the superconducting gap anisotropy [61,62]. For points 2 and 3 which locate close to the boundary of the main WDR, their $\alpha$ values are larger than those of points 1 and 4 [brown squares in Fig. 3(h)]. Intriguingly, we find that $\Lambda_0^{-1}$ (blue circles) tracks the local $T_c$ (yellow circles) closely for these points, which suggests a strong correlation between the two quantities.

During the scanning process, the nano-SQUID probe moves parallel to the film surface with a constant scan height of ~ 0.8 μm. Hence, the geometry factors in Eq. (1), i.e., $a$, $z$ and $z_0$ maintain constant values. In this regard, the value of $\Lambda^{-1}$ for any point in the scan area of Fig. 1(c) (denoted as R1) with respect to point $i$ ($i$ = 1-4) is equal to the ratio of their $\chi'$ values. This allows us to transform the $\chi'$ image [Fig. 1(c)] into the $\Lambda^{-1}$ image, as displayed in the left panel of Fig. 4(a) (see details in Supplementary Materials S6). Using this procedure, we have obtained the $\Lambda^{-1}$ images from three other scan areas (denoted as R2-R4), which all exhibit spatial variation on the micron scale [see the left panels in Figs. 4(b-d)]. The corresponding mappings of the local $T_c$ is provided in the right panels of Figs. 4(a-d), showing clear spatial correlation with the $\Lambda^{-1}$ images.

The $T_c$ versus $\Lambda^{-1}(T_{\text{base}})$ data extracted from different random areas across the sample exhibit a continuous trend [Fig. 4(e)]. For all the points in these areas, higher base-temperature superfluid density corresponds to a higher $T_c$. In particular, for R1-R2 with $T_c > 8$ K, the data (blue markers) appears to obey a linear relation with an offset, $T_c = T_0 + \alpha\Lambda^{-1}$ with $T_0 = 7.29$ K and $\alpha = 1.75 \times 10^3$ K·µm (the red dashed line). Whereas for R3-R4 with $T_c < 8$ K, the data (orange and red markers) suggest a sublinear relation between $T_c$ and $\Lambda^{-1}$ that can be captured by $T_c = \gamma\sqrt{\Lambda^{-1}}$ with $\gamma = 405$ K·µm$^{1/2}$ (the blue dashed line). To check whether the $T_c$ - $\Lambda^{-1}(T_{\text{base}})$ relationship is robust, we have extracted the $T_c$ versus $\Lambda^{-1}(T_{\text{base}})$ data from two scan areas (R5-R6) of the second NSNO film with $T_c$ ranging from 11.2-12.6 K (Supplementary Materials S8). As shown in the inset of Fig. 4(e), the data of R5 (yellow markers) and R6 (cyan markers) are in line with the extrapolated linear relation (the red dashed line) of R1-R2 [see also the enlarged view in Fig. S10(e)]. Moreover, data from STO capped $R_{0.8}$Sr$_{0.2}$NiO$_2$ ($R$ = La, Pr and Nd) measured by the mutual inductance technique [39] also follow a similar trend (black diamonds). Here, the $\Lambda^{-1}$ values of $R_{0.8}$Sr$_{0.2}$NiO$_2$ and R5-R6 are rescaled by constant factors of 0.15 and 2.9, respectively. The former may be related to the presence of STO capping layer in their samples, which has been found to influence the electronic structure of NSNO [21,63] and thereby potentially affect the superfluid density. Whereas for the two uncapped films studied here, the difference in $\Lambda^{-1}$ values is mainly attributed to the difference in their dead-layer thicknesses ($d_{\text{dead}}$) [64]. With the assumption of a negligible $d_{\text{dead}}$ in the first sample, the rescaling factor of 2.9 indicates $d_{\text{dead}} \approx 4.8$ nm for the second NSNO film, which is close to the reported value of 4.6 nm for some uncapped NSNO films [65].

In order to obtain the relation between $T_c$ and zero-temperature superfluid density $\Lambda_0^{-1} = \Lambda^{-1}(T \to 0 \text{ K})$, we extrapolate $\Lambda^{-1}(T)$ at finite temperatures to $T = 0$ K using both nodeless-gap and nodal-gap models, which have been widely used to analysis the temperature dependence of superfluid density for high-$T_c$ superconductors [61,66,67] including IL nickelates [39] (see details in Supplementary Materials S7). Due to the temperature limitations of the scanning SQUID device, it is difficult to determine which model is better to fitting the raw $\Lambda^{-1}(T)$ data [Fig. S8(a)]. Nevertheless, we find that the $T_c(\Lambda_0^{-1})$ data obeys the similar scaling relations between $T_c$ and $\Lambda^{-1}(T_{\text{base}})$, regardless of the gap model employed [see Fig. S8(d) and Fig. S9(c) accounting for the errors in extrapolating $\Lambda_0^{-1}$]. We note that NSNO may host multi-band or multi-gap superconductivity, which could result in a kink in $\Lambda^{-1}(T)$ when the interband coupling is weak [68,69]. However, such a kink is absent in $\Lambda^{-1}(T)$ of our samples and that of the NSNO films studied previously [34,39].

Remarkably, the scaling between $T_c$ and the superfluid density we observe is conspicuously similar to that observed in overdoped LSCO films using the bulk mutual inductance technique [13]. By rescaling the values of $T_c$ and $\Lambda^{-1}$ for LSCO [black circles in Fig, 4(e)], the two datasets overlap almost exactly, except for the high-$T_c$ region where the slope of LSCO is larger than that of NSNO. It should be noted that the superfluid density data of LSCO were accumulated from more than 2000 films prepared over about 12 years [70]. In contrast, the method of susceptometry imaging combined with spatial statistics, which proposed for the first time in this work, enables the acquisition of continuous and dense $T_c(\Lambda^{-1})$ data on a single sample containing inhomogeneity. Such a high-efficiency methodology also avoids errors caused by the variations in growth condition from one sample to another in the bulk measurements [71,72]. A similar method for extracting the $T_c$ - $\Lambda^{-1}$ relationship from a single sample has been recently applied to iron-based superconductors [16]. On the other hand, we noticed that the variation in $T_c$ of NSNO (from 5.5 K to 12.6 K) is less than that of LSCO (from 4 K to 45 K) due to the limited maximum $T_c$ of NSNO. Future superfluid density measurements on other nickelate compounds with maximum $T_c$ exceeding 20 K, e.g., Sm$_{1-x-y-z}$Ca$_x$Sr$_y$Eu$_z$NiO$_2$ [73] and Sr- or Pr-doped La$_3$Ni$_2$O$_7$ [74–76] will help examine whether the scaling relation extends to $T_c > 12.5$ K.

The linear relationship between $T_c$ and the superfluid density has been considered as an empirical evidence that $T_c$ is dominated by the phase stiffness [70,77,78], in contrast to the standard BCS paradigm that $T_c$ is dominated by the pairing strength. The long-range phase coherence is destroyed at $T_c$ owing to stronger thermal phase fluctuations with decreasing superfluid density. Thus, the observed linear scaling between $T_c$ and superfluid density in R1-R2 may also suggest significant thermal phase fluctuations in NSNO. This is supported by the rapid drop of $\Lambda^{-1}(T)$ above $T_c$ [see Fig. 3(g) and also the mutual inductance measurements of IL nickelates [39], and by the crossover from 3 to 1 in the power-law exponent of the current-voltage characteristic curves [79–81]. Furthermore, a recent THz pump-probe study on NSNO also points out substantial superconducting fluctuations near $T_c$ [82]. For R3 and R4 with $T_c < 8$ K, the sublinear relation $T_c = \gamma\sqrt{\Lambda^{-1}}$ can be understood in terms of quantum phase fluctuations within the (3+1)D-$XY$ universality class with $z = 1$ as $T_c \to 0$ K [11], which has been adopted to understand the similar relation for overdoped LSCO with $T_c < 12$ K [12].

On the other hand, several theoretical studies have shown that the $T_c$ versus superfluid density relation in overdoped LSCO can be reproduced in the calculations on the $d$-wave BCS theory by considering the effects of disorder scattering [83–85]. However, their calculations either depend critically on the specific band structure of LSCO, or involve the anisotropic scattering of apical oxygen vacancies that is absent for the infinite-layer nickelates. Whether these disorder scenarios could be extended to explain the quantitative relation between $T_c$ and superfluid density in NSNO remains unclear. Regardless, the similarity in the $T_c$ versus superfluid density relation between cuprate superconductors and IL nickelates suggests a close underlying principle governing the superconductivity of these two high-$T_c$ families, despite their distinct electronic structures and pairing symmetries.

In summary, we perform local susceptometry on the infinite-layer NSNO/STO films and establish a quantitative relation between local superfluid density and $T_c$. We find that their relation is strikingly similar to that found in overdoped cuprate superconductors. Our findings suggest the unconventional pairing beyond BCS paradigm in infinite-layer nickelates and its close ties with cuprates.

**Acknowledgements**

We would like to acknowledge support by National Natural Science Foundation of China (Grant No. 12150003 and 12274061), Shanghai Municipal Science and Technology Major Project (Grant No. 2019SHZDZX01), National Key R&D Program of China (Grant No. 2021YFA1400100 and 2023YFA1406301), and China Postdoctoral Science Foundation (Grant No. 2025M773424).

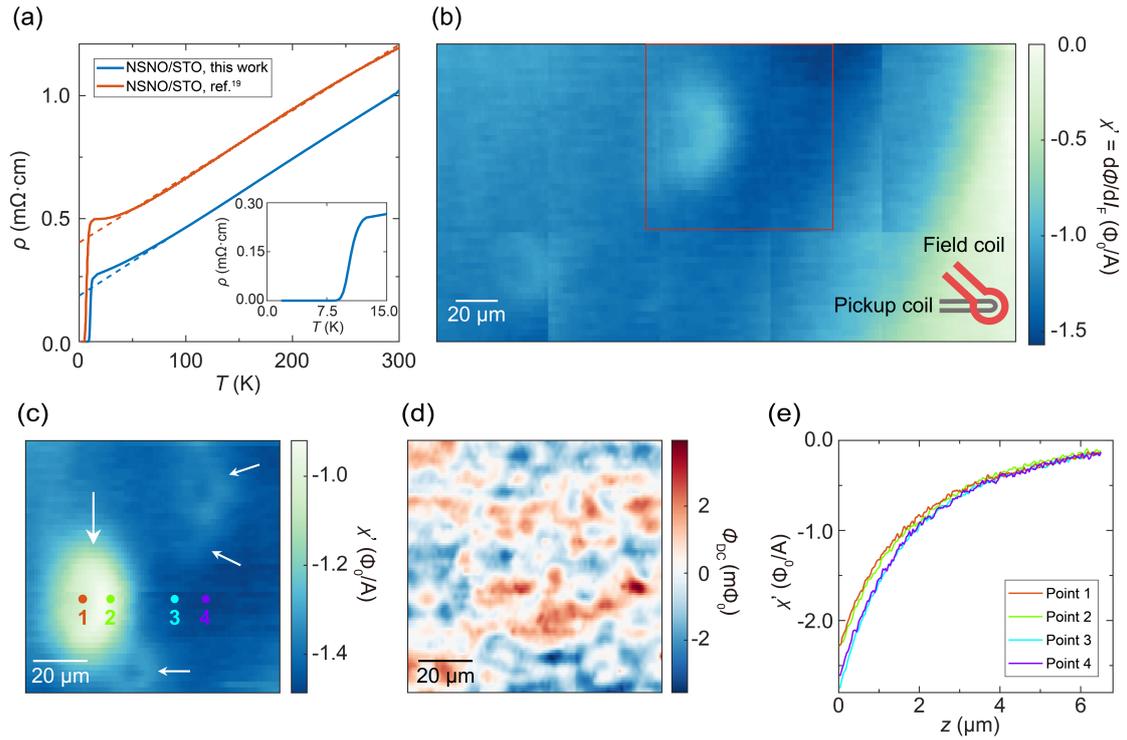

**FIG. 1 Characterization of superconducting properties of the $Nd_{0.8}Sr_{0.2}NiO_2$ (NSNO) film.** All data are acquired at $T_{base} = 5$ K except specified otherwise. (a) Temperature-dependent resistivities of NSNO films. (b) In-phase susceptibility imaging of a 280 × 150 μm² region. Negative $\chi'$ (blue color) indicates diamagnetism. Multiple weakly diamagnetic rings (WDRs) are visible. The red square depicts the scan area in panel (c) and Fig. 2. The inset shows the configuration and orientation of the pickup loop (gray) and the field coil (red) of the SQUID sensor. (c) In-phase susceptibility imaging of the main WDR (indicated by the large arrow) in panel (b). Smaller WDRs (indicated by small arrows) are visible. (d) DC magnetic flux imaging of the same region as in (c). (e) Susceptibility approach curves $\chi'(z)$ of the selected points marked in panel (c).

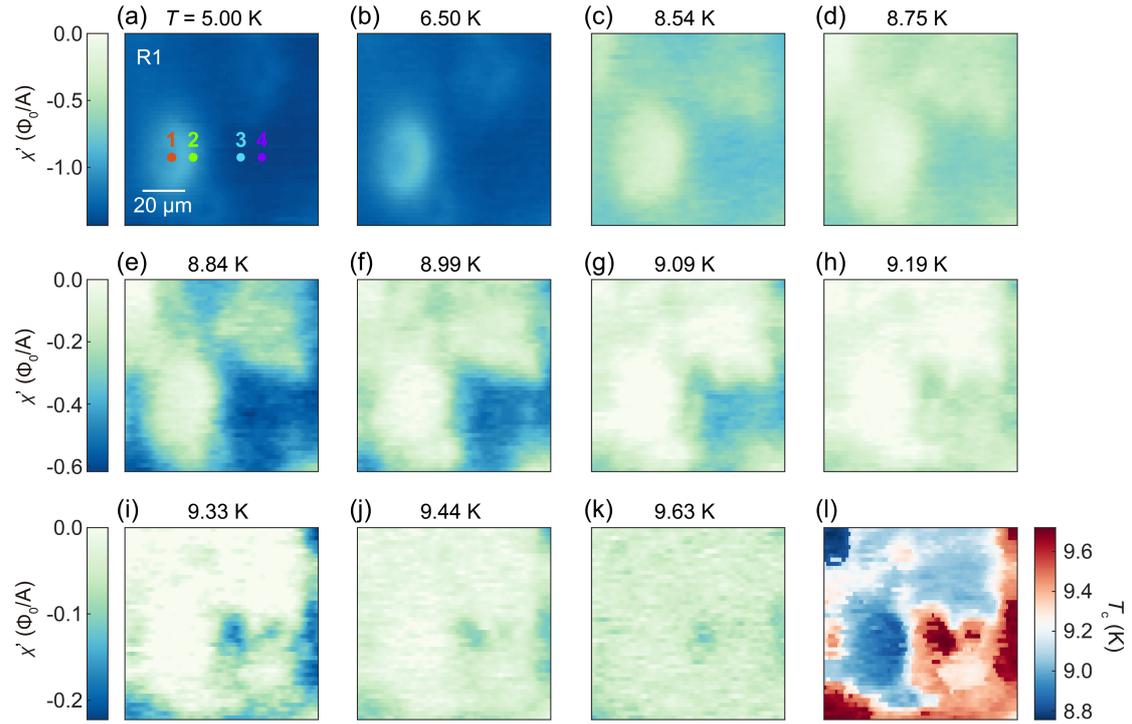

**FIG. 2 Temperature evolution of the superconducting diamagnetism and the map of local superconducting transition temperature.** The scan area, labeled as R1, is the same as in Fig. 1(c). For (a)-(k), every panel in a row shares the same color scale. At $T <$ bulk $T_c = 8.6$ K, the scan area exhibits diamagnetism with significant inhomogeneity [(a)-(c)]. At bulk $T_c < T \leq$ 9.63 K, WDRs and adjacent regions lose diamagnetism quickly with increasing temperature. Beyond 9.63 K, the diamagnetism of the full scan area vanishes (not shown in the figure). Panel (l) is the map of local $T_c$, which is determined as the onset temperature of superconducting diamagnetism.

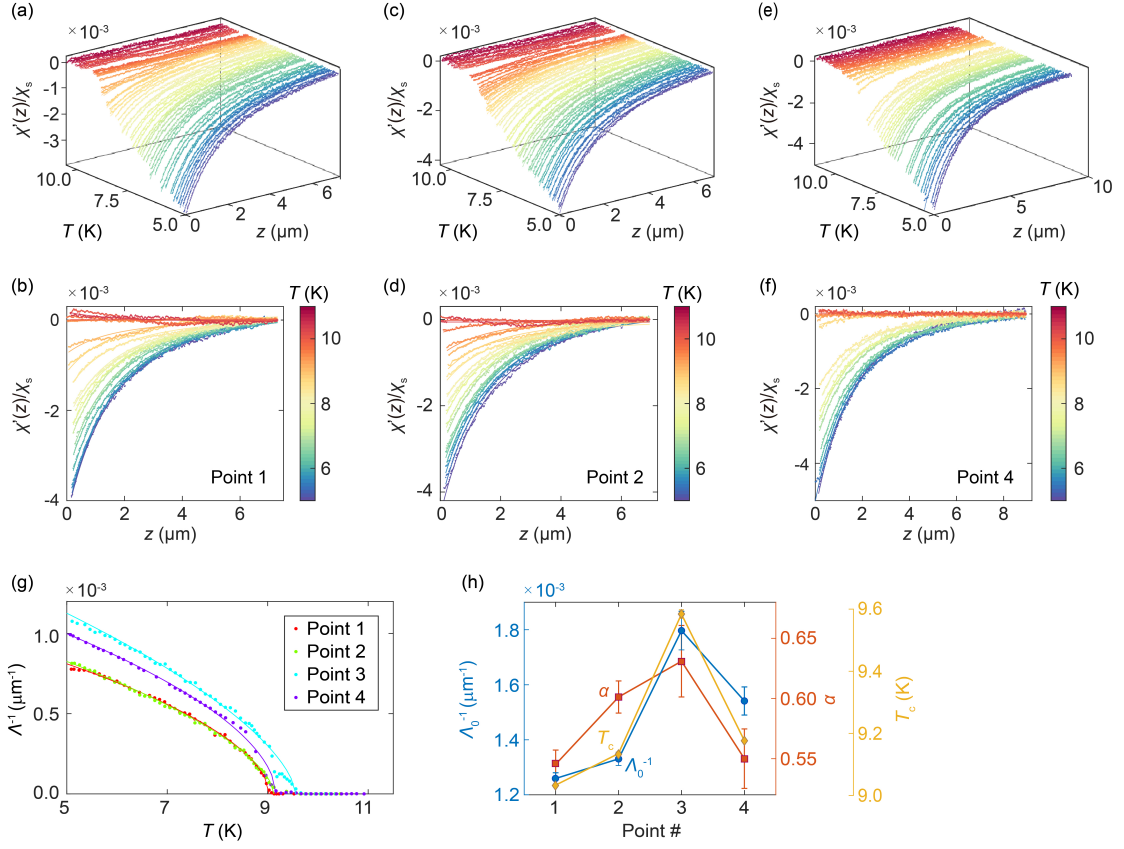

**FIG. 3 Temperature evolution of susceptibility approach curves and superfluid density at different points.** Panels (a)-(f) show the susceptibility approach data measured at the points labelled in Fig. 1(c): (a)-(b) correspond to point 1, (c)-(d) point 2, and (e)-(f) point 4. The color of data points represents the temperature. The temperature increment is 0.1 K in (a), (c), (e) and 0.3 K in (b), (d), (f). The solid lines are fits of the data using Eq. (1). (g) Temperature-dependent $\Lambda^{-1}(T)$. Dots represent experimental data, and the curves are fittings of the experimental data to Eq. (2). The error bars of the fitting are smaller than the size of data points. (h) Fitting parameters of points 1-4. Left axis (blue): zero-temperature $\Lambda_0^{-1}$, which is proportional to the superfluid density at $T = 0$ K; right axis (inner, brown): $\alpha$; right axis (outer, yellow): $T_c$.

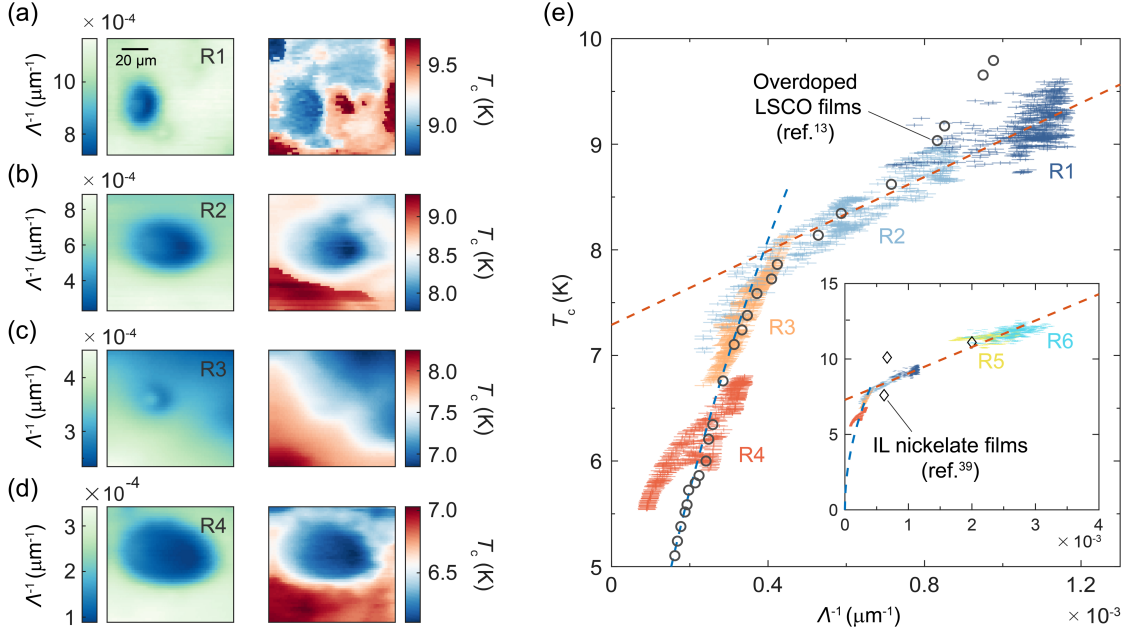

**FIG. 4 Correlation between superfluid density and local superconducting transition temperature in NSNO.** (a-d) Mappings of the inverse of Pearl length $\Lambda^{-1}$ at base temperature $T_{\text{base}} = 5$ K (left column), and the local critical temperature $T_c$ (right column) for regions R1-R4. (e) Scaling of $T_c$ with $\Lambda^{-1}$ extracted from (a-d). The color of the markers represents the scan area from which the data are extracted. The length and the width of markers represent the uncertainties in determining local $T_c$ and $\Lambda^{-1}$, respectively. Data points with $T_c > 8$ K can be fitted with a linear relation $T_c = T_0 + \alpha \Lambda^{-1}$ where $T_0 = 7.29$ K, $\alpha = 1.75 \times 10^3$ K·μm (the red dashed line), and those with $T_c < 8$ K can be fitted with a parabolic relation $T_c = \gamma\sqrt{\Lambda^{-1}}$ where $\gamma = 405$ K·μm$^{1/2}$ (the blue dashed line). For reference, we include the data of overdoped LSCO [13], where the values of $T_c$ and $\Lambda^{-1}$ have been rescaled (black circles). The inset provides the $T_c$ versus $\Lambda^{-1}$ data extracted from two scan regions (R5-R6) of the second NSNO film with higher $T_c$ (Supplementary Material S8), alongside data from STO capped $R_{0.8}\text{Sr}_{0.2}\text{NiO}_2$ films measured by the mutual inductance technique [39] (black diamonds). Here, the $\Lambda^{-1}$ values for R5-R6 and $R_{0.8}\text{Sr}_{0.2}\text{NiO}_2$ are rescaled by constant factors of 2.9 and 0.15, respectively.